\documentclass[preprint,showpacs,amsmath,amssymb]{revtex4}
\usepackage{graphicx} 
\usepackage{dcolumn}
\usepackage{bm}

\begin{document}
\title{Beating of the 
oscillations in  the transport coefficients of
a one-dimensionally
periodically modulated two-dimensional electron gas
in  the presence of spin-orbit interaction}
\author{X. F. Wang}
\email{xuefeng@alcor.concordia.ca}
\author{P. Vasilopoulos}
\email{takis@alcor.concordia.ca}
\address{Department of Physics, Concordia University\\
1455 de Maisonneuve  Ouest, Montr\'{e}al, Qu\'{e}bec, Canada, H3G 1M8}
\author{F.M. Peeters}
\email{peeters@uia.ac.be}
\address{Departement Fysica, Universiteit Antwerpen\\
(Campus Drie Eiken) Universiteitsplein 1,
 B-2610 Antwerpen, Belgium}
\begin{abstract}
Transport properties of a two-dimensional electron gas (2DEG)
are studied in the presence of a perpendicular magnetic field $B$,
 of a {\it weak} one-dimensional (1D) periodic potential  modulation,
 and of the  spin-orbit
interaction (SOI) described only by the Rashba term. In the
absence of the  modulation the SOI mixes the spin-up and spin-down
states of neighboring Landau levels into two new, unequally spaced
energy branches. The levels of these branches broaden into bands
in the presence of the  modulation and their bandwidths oscillate
with the field $B$. Evaluated at the Fermi energy, the $n$-th level
bandwidth of each series  has a minimum or vanishes at different
values of the field $B$. In contrast with the 1D-modulated 2DEG
without SOI, for which only one flat-band condition applies, here
there are two
flat-band conditions that can  change considerably as a function
of the SOI strength $\alpha$
and accordingly influence the transport coefficients
of the 2DEG.
The phase and amplitude of the Weiss and Shubnikov-de Haas (SdH) oscillations
depend on the strength $\alpha$.
For small values of $\alpha$ both oscillations show beating patterns. Those of  
the former are due to the independently oscillating bandwidths whereas those of the latter
are due to modifications of the density of states, exhibit an even-odd filling factor transition,
and are nearly independent of the modulation strength.
For strong values of $\alpha$ the SdH oscillations are split in two.
\end{abstract}
\date{\today}
\maketitle

\section{ INTRODUCTION}
 The magnetotransport of the 2DEG,
subjected to periodic potential modulations, has attracted
considerable experimental \cite{wei} and theoretical
\cite{ger,pee} attention during the last two decades. For
one-dimensional (1D) modulations novel oscillations of the
magnetoresistivity tensor $\rho_{\mu\nu}$ have been observed, at
low magnetic fields $B$, distinctly different in period and
temperature dependence from the usual Shubnikov-de Haas (SdH) ones
observed at higher $B$. These novel oscillations, referred to as
the Weiss  oscillations, reflect the commensurability between two
length scales: the cyclotron diameter at the Fermi level $2R_c =
2\sqrt{2\pi n_e} \ell_c^2$, with $n_e$   the electron density and
$\ell_c$ the magnetic length, and $a$ the period  of the potential
modulation.

The emerging field of spintronics brought into
the fore
the importance
of spin-orbit interaction (SOI)  in a variety of situations. It is important
in the development of spin-based transistors
\cite{dat}, possibly in future quantum computations \cite{kik},
in an unexpected metal-to-insulator
transition in 2D \cite{tut} hole gas, in spin-resolved ballistic transport \cite{lu},
in Aharonov-Casher  experiments \cite{mor},  in  spin-galvanic  \cite{gan} and spin valve  \cite{mir} effects, in the spin-Hall effect \cite{sin}, etc.
The effect is important
in inversely asymmetric bulk semiconductor  crystals, due to the internal crystal
field, as well in asymmetrically confined semiconductor heterostructures. In the former case
the contributions to the spin splitting in the conduction band  vary as a $\sim k^3$ term
and dominate in {\it wide-gap}
structures \cite{dre} whereas in the latter vary as a
$\sim k$ term, referred to as the Rashba  term, and dominate in {\it narrow-gap}
structures \cite{ras}.  The latter was  confirmed by experiments that showed
a zero-magnetic-field spin splitting for carriers with finite momentum in a
modulation-doped GaAs/AlGaAs heterojunction  \cite{sto}
as well as by magnetotransport  measurements in a 2D hole
system \cite{luo}. The explanation proposed by Bychkov and Rashba \cite{byc} employed the
Rashba spin-orbit Hamiltonian, in which the spin of finite-momentum
electrons feels a magnetic field perpendicular to the electron momentum in the inversion plane. A detailed
account of magnetotransport of the 2DEG in the presence of SOI but absence of modulations
appeared recently \cite{wan}.

Given the importance the SOI has acquired, one  question that arises concerns its influence on
magnetotransport properties of a 2DEG in  the presence of periodic potential modulations. So far we are
aware of only the brief, {\it classical} study of Ref. \cite{mag}. Since some effects of the modulations can
be explained only quantum mechanically  \cite{pee}, it is of interest to reexamine the problem {\it quantum
mechanically}. This is the subject of this paper. We will consider only weak 1D modulations and make use of
our experience with them \cite{pee} and with the unmodulated 2DEG in the presence of SOI \cite{wan}. The
main qualitative findings are as follows. The levels of the $+$ and $-$, unequally spaced energy
branches, due to the SOI when the modulation is absent, broaden into bands when the  modulation is present
and their bandwidths oscillate with the field $B$. Evaluated at the Fermi energy, these bandwidths vanish
at different values of the  field $B$ and modify considerably
the flat-band condition and the transport coefficients as a function of the SOI strength $\alpha$.
As a result , the phase and amplitude of the commensurability and SdH oscillations 
change when  $\alpha$ is varied.
For small values of $\alpha$ the former show a beating pattern while for strong values of $\alpha$ the 
latter are split in two.

In the next section we present the one-electron eigenfunctions and
eigenvalues. Analytical results for the conductivities are given in Sec. III and numerical results
in Sec. IV. The last section contains a summary and
concluding remarks.

\section{ Eigenfunctions and eigenvalues}

\subsection{A 2DEG in the presence of SOI and absence of  potential modulation}

We consider a 2DEG in the $(x-y)$ plane and a magnetic field along the $z$
direction. In the Landau gauge ${\bf A}=(0,Bx,0)$ the one-electron
Hamiltonian including the Rashba term reads
\begin{equation}
H_{0}=\frac{({\bf p}+e{\bf A})^{2}}{2m^{\ast }}+\frac{\alpha
}{\hbar }\left[ {\mbox{\boldmath $\sigma$} }\times ({\bf p}+e{\bf
A})\right] _{z}+\frac{1}{2}g\mu _{B}B\sigma _{z}, \label{Ham}
\end{equation}
where ${\bf p}$ is the momentum operator of the electrons,
$m^{\ast }$ is the effective electron mass,$\ g$ the Zeeman
factor, $\mu _{B}$ the Bohr magneton,
${\mbox{\boldmath $\sigma$} }=(\sigma_{x},\sigma _{y},\sigma _{z})$
the Pauli spin matrices, and
$\alpha $ the strength of the SOI or Rashba parameter.

Using the Landau wave functions without SOI as a basis, we can express the
new eigenfunction in the form:
\begin{equation}
\Psi _{k_{y}}({\bf r}) = e^{ik_{y}y}\sum_{n=0}^\infty\phi
_{n}(x+x_{c})\left( \tiny
\begin{array}{c}
C_{n}^{+}\\
C_{n}^{-}
\end{array}
\right) /\sqrt{L_{y}}. 
\label{wav}
\end{equation}
Here $\phi
_{n}(x)=e^{-x^{2}/2l_{c}^{2}}H_{n}(x/l_{c})/(
\sqrt{\pi }2^{n}n!l_{c})^{1/2}$ is the  harmonic oscillator function, $\omega
_{c}=eB/m^{\ast }$ the cyclotron frequency, $l_{c}=(\hbar /m^{\ast }\omega
_{c})^{1/2}$ the magnetic length, and the cyclotron orbit is centered at
$x_{c}=l_{c}^{2}k_{y}$, $n$ the Landau-level index, and
$|\sigma \rangle$
the electron spin written as the row vector
$\langle\sigma| =\left( 1, \  0\right) $
if it's pointing up and $\left(0, \ 1\right) $ if it's pointing down.

Using these wave functions and Eq. (1) the eigenvalue problem
$H_0\Psi =E\Psi $ leads to an infinite system of  equations that
can be solved exactly after decomposing it into independent systems
of one or two  equations \cite{wan}. The resulting eigenstates are
labelled by a new quantum number $s$ for  the energy instead of $n$.
For $s=0$  there is one level, the same as the lowest Landau
level without SOI, with energy
\begin{equation}
E_{0}^{+}=E_{0}=\hbar \omega _{c}/2-g\mu _{B}B/2
\end{equation}
and wave function
\begin{equation}
\Psi _{0}^{+}(k_{y})=
e^{ik_{y}y}
\phi _{0}(x+x_{c})
\left(\tiny
\begin{array}{c}
0 \\
1
\end{array}
\right)/\sqrt{L_{y}}.
\end{equation}

For $s=1,2,3,\cdots $, there are two branches of levels, denoted by $+$ and $-$, with energies
\begin{equation}
E_{s}^{\pm }=s\hbar \omega _{c}\pm \lbrack E_{0}^{2}+2s\alpha
^{2}/l_{c}^{2}]^{1/2}.  \label{eig}
\end{equation}
The $+$ branch is described by the wave function
\begin{equation}
\Psi _{s}^{+}(k_{y})=\frac{e^{ik_{y}y}}{\sqrt{L_{y}{\cal A}_{s}}}\left(
\begin{array}{r}
{\cal D}_{s}\phi _{s-1}(x+x_{c})
\\
\phi _{s}(x+x_{c})
\end{array}
\right),
\end{equation}
and the $-$ one by
\begin{equation}
\Psi _{s}^{-}(k_{y})=\frac{e^{ik_{y}y}}{\sqrt{L_{y}{\cal A}_{s}}}\left(
\begin{array}{r}
\phi _{s-1}(x+x_{c})
\\
-{\cal D}_{s}\phi _{s}(x+x_{c})
\end{array}
\right) \text{,}
\end{equation}
where ${\cal A}_{s}=1+{\cal D}_{s}^{2}$ and
${\cal D}_{s}=(\sqrt{2s}\alpha /l_{c})/[E_{0}+\sqrt{E_{0}^{2}+2s\alpha
^{2}/l_{c}^{2}}]$.

\subsection{A 2DEG in the presence of SOI and  of a 1D potential modulation}

In the presence of a 1D periodic electric modulation, we consider the Hamiltonian
\begin{equation}
H= H_{0}+V_{0}\cos (Kx),
\end{equation}
with $K=2\pi /a$ and $a$ the modulation  period. For weak modulations
the energy correction due to the term $V_{0}\cos (Kx) $ is  evaluated by first-order   perturbation theory.
The results for the two branches are
\begin{equation}
E_{s}^{+}=s\hbar \omega _{c}+[E_{0}^{2}+2s\alpha
^{2}/l_{c}^{2}]^{1/2}+V_{0}e^{-u/2}\cos (Kx_{c})[{\cal D}%
_{s}^{2}L_{s-1}(u)+L_{s}(u)]/{\cal A}_{s}; \  s=0,1,\cdots
\end{equation}
\begin{equation}
 E_{s}^{-}=s\hbar \omega _{c}-[E_{0}^{2}+2s\alpha
^{2}/l_{c}^{2}]^{1/2}+V_{0}e^{-u/2}\cos (Kx_{c})[L_{s-1}(u)+{\cal D}%
_{s}^{2}L_{s}(u)]/{\cal A}_{s}; \  s=1,2,\cdots
\end{equation}
where $u=2\pi ^{2}l_{c}^{2}/a^{2}=K^{2}l_{c}^{2}/2$ and $
x_{c}=k_{y}l_{c}^{2}$. $L_{s}(u)$ is the Laguerre polynomial and
for $s=0$ Eq. (9) reduces to Eq. (3) as modified by the
perturbation correction. The width of the broadened levels of the
two branches
is given by twice the absolute value of the last  term in Eqs. (9) and (10) without the
$\cos(K\ell^2_ck_y)$ factor and is denoted by $2|\Delta_s^{\pm }|$.  $\Delta_s^{\pm }$
can be written in the compact form
\begin{equation}
\Delta_s^{\pm }=V_0e^{-u/2}
[L_{s-1/2\pm 1/2}+{\cal D}_s^2 L_{s-1/2\mp 1/2}]
/{\cal A}_s,
\label{in1}
\end{equation}
with  the upper signs pertaining to the $+$ branch and the lower ones to the $-$ branch; obviously $\Delta_s^{\pm }$ is not the same for the two branches. In contrast, without SOI we have only a single branch and a single bandwidth
\cite{pee} and the eigenvalues are  given, when the Zeeman term is neglected, by
\begin{equation}
E_n=(n+1/2)\hbar \omega _{c}+V_{0}e^{-u/2}\cos (Kx_{c})L_{n}(u)
\end{equation}
with $n$ the Landau-level index. This has consequences that will be detailed below.

As in the absence of SOI, the presence of the modulation broadens the
discrete levels into bands. An important difference with the situation in
which the modulation is absent is that the diagonal matrix elements of the
velocity operator now do not vanish. Using $v_{y}^{\pm} =
(1/\hbar )\partial E_{s}^{\pm}(k_{y})/\partial k_{y}$ their values are
\begin{equation}
v_{y}^{+}
=-2V_{0}u e^{-u/2}[{\cal D}
_{s}^{2}L_{s-1}(u)+L_{s}(u)]\sin (Kx_{c})/(\hbar K{\cal A}_{s}),
\end{equation}
\begin{equation}
v_{y}^{-}   =-2V_{0}u e^{-u/2}[L_{s-1}(u)+{\cal D}
_{s}^{2}L_{s}(u)]\sin (Kx_{c})/(\hbar K{\cal A}_{s}).
\end{equation}
These non vanishing values lead to a non vanishing diffusive
conductivity whereas  in the absence of the modulation this
conductivity vanishes whether the SOI is present  or not \cite{wan}.
Compared to the case  without SOI, we have
two contributions, one
from Eq. (9) and one from Eq. (10), while for $\alpha=0$ we
have only one value given by
\begin{equation}
v_{y}=-(2V_{0}/\hbar K) u e^{-u/2}L_{n}(u)\sin (Kx_{c}).
\end{equation}
As a function of the magnetic field $B$, these $v_y^\pm$ contributions do not oscillate
in phase due to the different dependence of the Laguerre polynomials on $B$.
This modifies mostly the diffusive conductivity in  the presence of the modulation
and will be detailed in the following sections.

Using the asymptotic expression of the Laguerre polynomials for large $s$,
we obtain $\Delta_s^{\pm } \propto \cos(2\sqrt{su}-\pi/4)$.
The  Landau level indices $s^+$ and $s^-$ of the corresponding branches at the Fermi energy
can be determined by the equations $E_{s^+}^+ \approx E_{s^-}^-$ and $n_e = (s^++s^-+1)/(2\pi l_c^2)$,
where $n_e$ is the electron density. Then from the argument of $\cos(2\sqrt{su}-\pi/4)$ we obtain the
flat-band conditions 
\begin{equation}
\sqrt{u}[\sqrt{\pi n_e} l_c \mp \alpha/(\sqrt{2}\hbar\omega_c
l_c)]=\pi(i-1/4)/2 \label{flatband}
\end{equation}
with the upper (lower) sign corresponding to the $+$ $(-)$ branch. Since the cyclotron radius
at the Fermi energy is $R_c^{\pm}=l_c\sqrt{2s^{\pm}+1}$, Eq. (16)
can be written as
$2R_c^{\pm}/a=i-1/4$ with $R_c^{\pm}=R_c^0\mp \alpha/\hbar\omega_c$ and $R_c^0$
the cyclotron radius without SOI
or $K(k_F\mp k_\alpha)l_c^2=\pi(i-1/4)$ with $k_F=\sqrt{2\pi n_e}$,
and $k_\alpha=\alpha m^\ast/\hbar^2$. The same result has been obtained
in Ref. \cite{mag} by  a purely classical treatment. The fact that now we have {\it two}
flat-band conditions, as opposed to {\it one} for $\alpha=0$, leads to oscillations with
{\it two} different frequencies and
consequently to beating patterns that will be shown in Sec. IV.
Explicitly, writing Eq. (16) again for $i\to i+1$ and subtracting the result from
Eq. (16), gives the periods in the $\pm$ branches as
$\Omega^+=ea/[2\hbar(k_F-k_\alpha)]$ and $\Omega^-=ea/[2\hbar(k_F+k_\alpha)]$.

\section{Conductivities}
For weak electric fields $E_{\nu }$, i.e., for linear responses, and weak
scattering potentials the expressions for the direct current (dc) conductivity
tensor $\sigma_{\mu \nu }$, in the one-electron approximation, reviewed in
Ref. \cite{vas}, reads $\sigma _{\mu \nu }=$ $\sigma _{\mu \nu }^{d}+\sigma _{\mu \nu
}^{nd}$ with $\mu ,\nu =x,y,z$. The terms $\sigma _{\mu \nu }^{d}$ and $\sigma _{\mu
\nu }^{nd}$\ stem from the diagonal and nondiagonal part of the density
operator $\widehat{\rho }$, respectively,
 in a given basis and $\langle J_{\mu }\rangle =Tr(\widehat{\rho }J_{\mu
})=\sigma _{\mu \nu }E_{\nu }$. In general, we have $\sigma _{\mu \nu }^{d}=$
$\sigma_{\mu \nu }^{dif}+\sigma _{\mu \nu }^{col}$.
The term $\sigma_{\mu \nu
}^{dif}$ describes the diffusive motion of electrons
and the term $\sigma _{\mu \nu }^{col}$ the collision
contributions or hopping. The former is given by
\begin{equation}
\sigma _{\mu\nu }^{dif}=\frac{\beta e^{2}}{S_{0}}\sum_{\zeta }
f(E_{s}^{\sigma} )[1-f(E_{s}^{\sigma} )]
\tau ^{\zeta }(E_{s}^{\sigma})
v_{\mu }^{\zeta }v_{\nu }^{\zeta },  \label{diff}
\end{equation}
where $\zeta \equiv (s,\sigma,k_{y})$ denotes the quantum numbers, $
v_{\mu }^{\zeta }=\langle \zeta |v_{\mu }|\zeta \rangle $ is the diagonal
element of the velocity operator $v_{\mu }$, and
$f(\varepsilon )$ the Fermi-Dirac function. Further,
$\tau ^{\zeta }(E_{s}^{\sigma})$ is
the relaxation time for elastic scattering,
$\beta=1/k_BT$, and $S_0$ is the area
of the system.

The term $\sigma _{\mu \nu }^{col}$ can be written  in the form
\begin{equation}
\sigma _{yy}^{col}=\frac{\beta e^{2}}{2S_{0}}\sum_{\zeta ,\zeta ^{\prime
}}\int_{-\infty }^{\infty }d\varepsilon \int_{-\infty }^{\infty
}d\varepsilon ^{\prime }\delta \lbrack \varepsilon -E_{s}^{\sigma
}(k_{x})]\delta \lbrack \varepsilon ^{\prime }-E_{s^{\prime }}^{\sigma
^{\prime }}(k_{x}^{\prime })]f(\varepsilon )[1-f(\varepsilon ^{\prime
})]W_{\zeta \zeta ^{\prime }}(\varepsilon ,\varepsilon ^{\prime })(y_{\zeta
}-y_{\zeta ^{\prime }})^{2},  \label{col}
\end{equation}
where $y_{\zeta}=\langle \zeta |y|\zeta \rangle $;
$W_{\zeta \zeta ^{\prime
}}(\varepsilon ,\varepsilon ^{\prime })$  is the transition rate.
 For  elastic scattering by dilute impurities, of density $N_I$,
 we have
\begin{equation}
W_{\zeta \zeta ^{\prime }}(\varepsilon ,\varepsilon ^{\prime })=\frac{2\pi
N_{I}}{\hbar S_{0}}\sum_{{\bf q}}|U({\bf q})|^{2}
|F_{\zeta \zeta^{\prime }}(u)|^{2}
\delta (\varepsilon -\varepsilon ^{\prime })
\delta_{k_{x},k_{x}^{\prime }-q_{x}},
\label{rat}
\end{equation}
where $u=l_{c}^{2}q^{2}/2$ and $q^2=q_{x}^{2}+q_{y}^{2}$.
$U({\bf q})=(e^{2}/2\epsilon _{0}\epsilon )/(q+k_s)$  is the
Fourier transform of the screened impurity potential with $\epsilon$ the static dielectric constant,
$\epsilon_0$ the dielectric permittivity, and $k_s$ the screening wave vector.

The diffusion contribution given by
Eq. (17) becomes
\begin{equation}
\sigma _{yy}^{dif}=\frac{e^2}{h}\frac{4\beta u^2\tau}{\pi K}
\sum_{s,\sigma}
\int_0^{a/2l_c^2}dk_y
(\Delta_s^{\sigma})^2
\sin^2(Kl_c^2k_y)
f(E_{n,k_y}^{\sigma})[1-f(E_{n,k_y}^{\sigma})]
\label{diff1}
\end{equation}
with $\Delta_s^{\sigma}=\Delta_s^{\pm }$ given by Eq. (11). The
related contribution $\sigma _{xx}^{dif}$ is zero since the
velocity $v_x$ vanishes.

For weak potential modulations we can neglect
Landau-level mixing, i.e., we can take $s^{\prime }=s$.
Then noting that $\sigma _{xx}^{col}=\sigma
_{yy}^{col}$, $\sum_{{\bf q}}=(S_{0}/2\pi )\int_{0}^{\infty
}qdq=(S_{0}/2\pi l_{c}^{2})\int_{0}^{\infty }du$, and $\sum_{k_{x}}=
(S_{0}/2\pi l_{c}^{2})$, the collisional contribution given by Eq. (18) takes the form
\begin{equation}
\sigma _{yy}^{col}=\frac{e^2}{\hbar}\frac{N_{I}\beta}{2A_0}
\sum_{s,\sigma, k_y}\int_{0}^{\infty }du\left| F_{ss}^{\sigma
}(u)\right| ^{2}u\int_{-\infty }^{\infty} d\varepsilon [\delta
(\varepsilon-E_{s}^{\sigma })]^{2}
f(\varepsilon)[1-f(\varepsilon)]
\big |U(\sqrt{2u/l_{c}^{2}})\big |^{2},
\end{equation}
where
\begin{eqnarray}
\left| F_{ss}^{-}(u)\right| ^{2} &=&\{L_{s-1}(u)+{\cal D}_{s}^{2}L_{s}(u)
\}^{2}e^{-u}/{\cal A}_{s}^{2}, \\*
\left| F_{ss}^{+}(u)\right| ^{2} &=&\{{\cal D}_{s}^{2}L_{s-1}(u)+L_{s}(u)
\}^{2}e^{-u}/{\cal A}_{s}^{2}.
\label{fact}
\end{eqnarray}

The exponential $e^{-u}$ favors small values of $u$. Assuming
$b=k_{s}^{2}l_{c}^{2}/2\gg u$
we may neglect the term
$2u/l_{c}^{2}$ in the expression for $U(\sqrt{2u/l_{c}^{2}})$ and define
$U_0=U(0)$. We then obtain
\begin{equation}
\sigma _{yy}^{col}=\frac{e^2}{h}\frac{N_{I}U_0^2\beta}{\pi
a\Gamma} \sum_{s,\sigma } \int_0^{a/2l_c^2}dk_y I_s^{\sigma}
f(\varepsilon)[1-f(\varepsilon)], \label{largeb}
\end{equation}
where
\begin{equation}
I_{s}^{\pm }=[(2s\pm 1){\cal D}_{s}^{4}-2s{\cal D}_{s}^{2}+2s\pm 1]/{\cal A}
_{s}^{2}.
\label{in2}
\end{equation}

The impurity density $N_{I}$  determines the Landau Level broadening
$\Gamma =W_{\zeta \zeta ^{\prime }}
(\varepsilon ,\varepsilon ^{\prime})/\hbar $.
Evaluating $W_{\zeta \zeta ^{\prime }}(\varepsilon ,\varepsilon^{\prime})/\hbar$
in the $u\to 0$ limit without taking into account the SOI, we obtain $N_{I}\approx 4\pi
[(2\epsilon \epsilon _{0}/e^{2}) ]^{2}\Gamma /\hbar$.

The Hall conductivity $\sigma _{xy}^{nd}$ is given by
\begin{equation}
\sigma _{xy}^{nd}=\frac{2i\hbar e^{2}}{S_{0}}\sum_{\zeta ,\zeta ^{\prime
}}
f(E_{\zeta } )[1-f(E_{\zeta }^{\prime })]
< \zeta\mid v_x\mid \zeta' >
<\zeta'\mid v_y\mid\zeta >
\frac{1-e^{\beta (E_{\zeta } -E_{\zeta }
^{\prime })}}{(E_{\zeta }-E_{\zeta }^{\prime })^{2}}, \ \ \zeta ^{\prime}\neq\zeta.
\label{hall}
\end{equation}

The resistivity tensor $\rho _{\mu \nu }$ is given in terms of the conductivity tensor $\sigma _{\mu \nu}$
upon  using the standard expressions $\rho _{xx}=\sigma _{yy}/S$, $\rho_{yy}=\sigma
_{xx}/S$, $\rho _{yx}=\rho _{xy}=-\sigma _{yx}/S$, where $S=\sigma
_{xx}\sigma _{yy}-\sigma _{xy}\sigma _{yx}$.

\section{Numerical results}
In this section we present numerical  results for the bandwidth
and the two conductivities given by Eqs. (20) and (23) for various
values of the SOI strength $\alpha$, of the modulation strength
$V_0$ and  period $a$, of the electron density $n_e$, and of the
temperature $T$. We measure $\alpha$ in units of
$\alpha_0=10^{-11}$ eVm, $n_e$ in  units of $n_0=10^{-11}/cm^2$,
and use the effective mass of InAs $m^*=0.05m_0$ with $m_0$ the
free-electron mass.

\begin{figure}[tpb]
\includegraphics*[width=90mm]{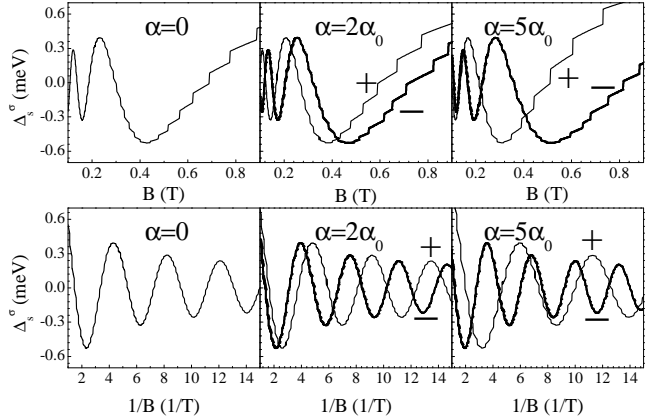}
\vspace{-3.5cm}
\caption{The  quantity $\Delta_s^\pm$ of Eq. (11)
vs magnetic field $B$
(upper panels) and vs inverse magnetic field $1/B$ (lower panels),
at the Fermi level, for different values of
the strength $\alpha$.
The modulation period is $a=3500$\AA \ and
the modulation strength $V_0=0.5$meV.}
\label{fig1}
\end{figure}

In Fig. \ref{fig1} we plot $\Delta^{\pm}_s$, given by Eq. (11) and  directly related to the bandwidth
2$|\Delta^{\pm}_s|$, at the Fermi level, as a function of the magnetic field $B$ in the upper panels
and as a function of the inverse magnetic field $1/B$ in the lower panels.
The other parameters are $a=3500$ \AA, $T=2$ K, $n_e=3n_0$, and
$V_0=0.5$ meV. We plot $\Delta_s^\pm$ and not 2$|\Delta^{\pm}_s|$
so that the oscillations are seen more clearly. Comparing the
$\alpha=0$ panel with the $\alpha\neq 0$ ones, we see clearly, for
$\alpha\neq 0$, the  contributions from the $+$ and $-$ branches.
The large-amplitude oscillations, for low $B$ in the upper panels
and for high $1/B$ in the lower panels, are the Weiss oscillations
whereas the step-like behavior on the right side of the upper
panels is due to the small-amplitude SdH ones. On the scale used
the latter are barely visible on the very left side in the lower
panels. The phase shift between the oscillations of
$\Delta_s^+$and $\Delta_s^-$ and their slightly different
frequencies described by Eq. (16) lead to the beating patterns of
the conductivities shown below.  For example, for
$\alpha=2\alpha_0$ the  oscillations of the bandwidth
$2|\Delta_s^\pm|$,
 given after Eq. (16), have a period
$\Omega^+=2.14$ T$^{-1}$ in the $+$
branch and a period $\Omega^-= 1.76$ T$^{-1}$ in the $-$ branch.

\begin{figure}[tpb]
\vspace{-4.5cm}
\includegraphics*[width=90mm]{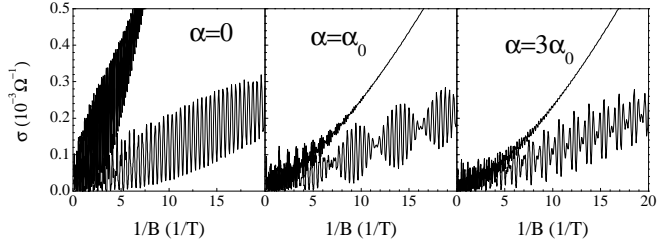}
\vspace{-4.5cm} \caption{Conductivities vs inverse  magnetic field
$B$ for different values of  $\alpha$ with $a=800$\AA, $T=1$K,
$n_e=3n_0$, and $V_0=0.3$ meV. The upper (lower) curves show the
collisional (diffusive) contribution.}
\label{fig2}
\end{figure}

In Fig. \ref{fig2} we plot the conductivities vs the inverse of the magnetic field $B$
for different values of $\alpha$
and a shorter  modulation period $a=800$ \AA.
The upper curve is the collisional conductivity,
given by Eq. (23), and the lower one the diffusive conductivity,
given by Eq. (20). Notice the absence of a beating pattern for $\alpha=0$
and its development for $\alpha\neq 0$. For finite $\alpha$, the longer-period
beating pattern of the Weiss oscillations is observed in the
diffusive curves and the shorter-period beating pattern of the SdH oscillations
\cite{wan} in the collisional curves. The reason is that
at low magnetic fields and low temperatures
the Weiss oscillations dominate the diffusive conductivity
while the SdH oscillations dominate the collisional conductivity.
In the former the energy correction due to the modulation, given by Eqs. (9)-(10),
enters mainly the square of $v^\pm_y$ and the argument of the Fermi function,
cf. Eqs. (17), (20), whereas in the latter it enters essentially only
through the argument of the Fermi function, cf. Eqs. (18), (24).

\begin{figure}[tpb]
\vspace{-2.5cm}
\includegraphics[width=90mm]{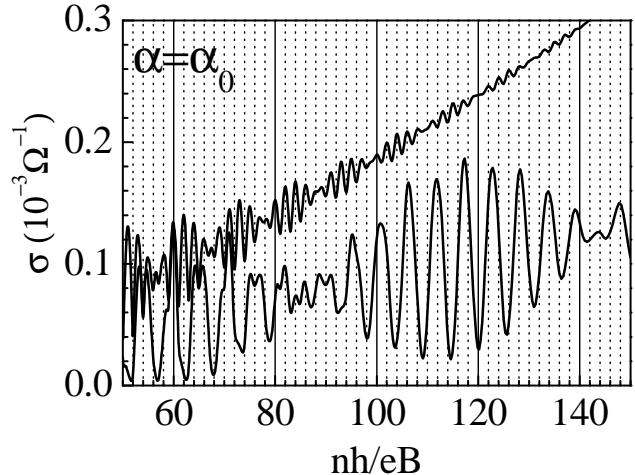}
\vspace{-1.5cm}
\caption{Conductivities vs filling factor $nh/eB$ for the panel $\alpha=\alpha_0$
of Fig. \ref{fig2}. The dashed vertical lines show the even filling factor values and the curves are marked as in Fig. \ref{fig2}.}
\label{fig3}
\end{figure}

\begin{figure}[tpb]
\vspace{-2.5cm}
\includegraphics[width=90mm]{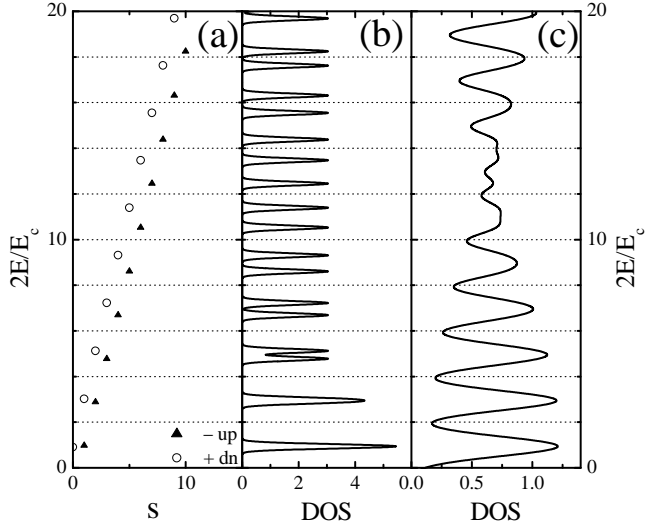}
\vspace{-2.5cm}
\caption{(a) Subband energy vs index $s$ in the absence of modulation. The DOS vs energy is shown in (b) for  subband broadening $\Gamma=0.1$ meV and in (c) for $\Gamma=0.5$ meV. When $E$ is the Fermi energy
the quantity $2E/E_c$ with $E_c=\hbar \omega_c$ is approximately the filling factor.}
\label{fig4}
\end{figure}

To see the oscillations shown in Fig. \ref{fig2} more clearly, we plot the
conductivities vs filling factor $ nh/eB$ in Fig. \ref{fig3}
for $\alpha=\alpha_0$ and $n_e=3n_0$.
As can be seen, the collisional conductivity (upper curve) shows a beating pattern of the SdH
oscillations resulting from the different Landau-level separations in the $+$ and $-$ spin branches.
The  index $s^\pm$ at the Fermi energy is expressed approximately as
$s^\pm=(\pi n_e \hbar\mp m^\ast\alpha\sqrt{2\pi n_e}/\hbar)/eB$. The resulting
period of the beating pattern, measured in units of inverse magnetic field,
is $2\hbar k_\alpha k_F/e$ or 0.85T$^{-1}$ in Fig. \ref{fig3}.
We notice that
a transition from  conductivity maxima
at {\it even} filling factors to conductivity maxima at {\it odd} filling factors
occurs between adjacent wraps of the SdH oscillations.
This can be understood by checking the DOS of the system. As shown in 
Fig. \ref{fig4},
when the subband broadening is comparable to the subband separation, a beating pattern appears
in the DOS, with SOI present and modulation absent, and each DOS peak corresponds
to one pair of spin levels.
Because the spin-up
and spin-down levels have different separations, there is one 
unpaired spin level at each node of the beating pattern. As a result, in one wrap of the
DOS oscillations there is an {\it even} number of levels below each pair and the DOS has
a peak at {\it odd} filling
factors, while in the next wrap there is an {\it odd} number of levels below each pair
and the DOS has a peak
at {\it even} filling factors. When the Fermi energy passes through the $\pm$ branches
and the DOS is as described above, the collisional
conductivity shows a beating pattern with an {\it even-odd} filling factor transition.
Although here the modulation is present, it is very weak and leaves the oscillations of the collisional
conductivity nearly intact. A complementary way of seeing how the beating pattern is formed,
is to plot separately
$\sigma^{col,-}$ and $\sigma^{col,+}$. Both contributions oscillate with slightly different
frequencies and their sum shows
the beating pattern of Fig. \ref{fig3}.
The period of this pattern, in units of
inverse of magnetic field, is
$ea/4\hbar k_\alpha$ or 4.63T$^{-1}$ in Fig. \ref{fig3}. A similar even-odd
filling factor transition was also observed for {\it strong}
modulations, which make the Landau levels overlap, in the absence of SOI and was explained by the behavior of the corresponding DOS \cite{shi}.
The diffusive conductivity (lower curve) shows mainly a beating pattern of the Weiss oscillations
 since here the SdH oscillations are very weak.

Above we observed a beating pattern in the SdH and Weiss oscillations occurring, respectively,
in the collisional and diffusive conductivities, vs filling factor when varying the magnetic
field at a fixed electron density.
If we vary the electron density  and fix the magnetic field $B$, the beating
pattern of the SdH oscillations holds because it corresponds to the Fermi energy passing through
the DOS with beating pattern. However, we do not observe a beating pattern in the Weiss oscillations.
This can be explained by Eq. (16), from where we see that, for fixed  $B$, the bandwidths of the
two series of spin levels
oscillate with the same frequency as a function of the electron density $n_e$ though with different phases.
For a system without the potential modulation, the diffusive conductivity disappears
and we observe only a beating pattern of the SdH oscillations in the collisional conductivity.

\begin{figure}[tpb]
\vspace{-4cm}
\includegraphics[width=90mm]{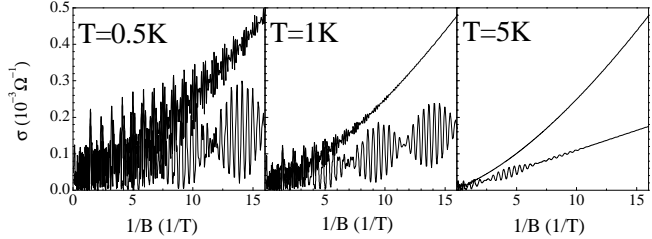}
\vspace{-2.5cm}
\caption{Conductivities vs inverse  magnetic field $B$ for different  temperatures
with $\alpha=\alpha_0$ and $n_e=3n_0$.
The upper (lower) curves show the
collisional (diffusive) contribution.}
\label{fig5}
\end{figure}

In Fig. \ref{fig5} we plot again the  conductivities vs the inverse of the
magnetic field $B$ for different values of  the temperature,
$\alpha=\alpha_0$, $n_e=3n_0$, and $V_0=0.3$ meV.
The two curves are marked as in Fig. \ref{fig2}. Notice that
beating pattern exists for all temperatures
but the oscillation amplitude decreases with increasing temperature
and nearly disappears at $T\approx 5$ K for the  density  and SOI strength  used.

\begin{figure}[tpb]
\vspace{-4.5cm}
\includegraphics*[width=90mm]{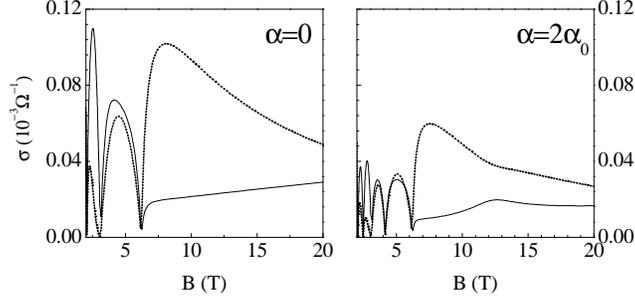}
\vspace{-2.5cm}
\caption{Conductivities vs magnetic field $B$ for two different values of $\alpha$.
The temperature is $T=1$K and the density $n_e=3n_0$.
The dotted (solid) curves show the collisional (diffusive) conductivity.}
\label{fig6}
\end{figure}

In Fig. \ref{fig6} we plot  the  conductivities vs  magnetic field $B$,
for rather strong values of $B$, and  different $\alpha$. The
temperature is $T=$1K.
The dotted (solid) curves show the collisional (diffusive) conductivity.
The SOI splits each Landau subband and reduces the DOS inside it.
As a result, a reduction in the oscillation amplitude and a splitting
of the (SdH) oscillations are observed in the $\alpha=2\alpha_0$ panel
compared with the $\alpha=0$ one.
For the high magnetic fields involved here, the period of the Weiss oscillations is very
long and both the diffusive and collisional conductivity curves show the SdH oscillations
with the same phase.

\begin{figure}[tpb]
\vspace{-4.5cm}
\includegraphics*[width=90mm]{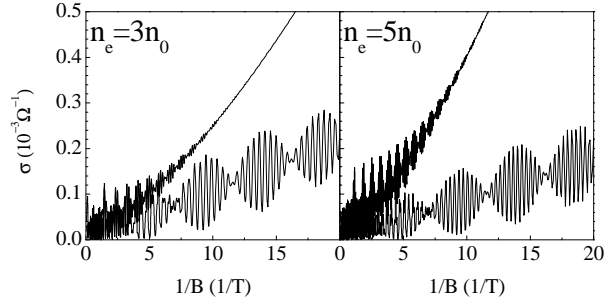}
\vspace{-2.5cm}
\caption{Conductivities vs inverse magnetic field $B$ for different densities, $\alpha=\alpha_0$,
and temperature $T=1$ K.
The upper (lower) curves show the collisional (diffusive) conductivity.}
\label{fig7}
\end{figure}

In Fig. \ref{fig7} we plot the conductivities vs magnetic field $B$ for different densities,
$\alpha=\alpha_0$,
and temperature $T=1$K.  Again the two curves are marked as in Fig. \ref{fig2}.
Notice how increasing the density
and thus changing the position of the Fermi level relative to those of the $+$ and $-$ branches closest
to it modifies the beating pattern.

\begin{figure}[tpb]
\vspace{-4cm}
\includegraphics*[width=90mm]{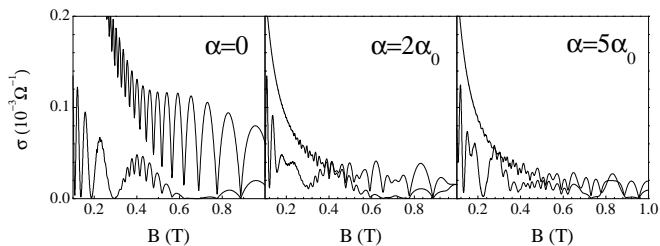}
\vspace{-4cm} 
\caption{Conductivities vs magnetic field $B$ for
different $\alpha$. The density is $n_e=3n_0$, the modulation
period $a=3500$\AA, and the temperature $T=2$ K.
The upper (lower) curves show the collisional (diffusive) conductivity.}
\label{fig8}
\end{figure}

In Fig. \ref{fig8} we plot  the  conductivities vs magnetic field $B$ for
different $\alpha$. The density is $n_e=3n_0$ and the temperature
$T=2$ K.
The upper (lower) curves are the collisional
(diffusive) contributions. 
The diffusive curve shows  mainly the Weiss oscillations  at low $B$ and at high $B$ the 
short-period SdH oscillations
in addition to the long-period Weiss oscillations. The collisional curve shows clearly the SdH oscillations for $\alpha=0$
and a beating pattern of the SdH oscillation for finite $\alpha$.

We now address the issue of the Hall conductivity $\sigma_{xy}^{nd}$.
In the absence of modulation and presence of SOI, it has
been evaluated in Ref. \cite{wan} for rather {\it strong} fields
$B\geq 1$ T and shows two series of quantum Hall plateaus, for
{\it strong} $\alpha$ ($\alpha\approx 10\alpha_0$), corresponding
to the two branches developed due the SOI.
The 1D modulation removes the $k_y$ degeneracy of the Landau levels $E_s$
and broadens them into bands with eigenvalues 
$E_{s,k_y}$.
From Eq. (26) we see that this may affect the Hall conductivity at weak magnetic fields when
the broadening $\Delta_s$ is comparable to the
energy $\hbar\omega_c$.
In the presence of modulation and absence of SOI, it
has been evaluated in Ref. \cite{pee} for {\it weak} fields
$B\leq 1$ T and shows very small-amplitude oscillations expressed mainly
through the energy difference between the $n$ and $n\pm 1$ Landau
levels. Here the interest is in
the region of {\it weak} fields $B\leq 1$ T for which the Weiss
oscillations appear. Despite the fact that $\Delta_s$ is comparable to  $\hbar\omega_c$, it exhibits again very small-amplitude
oscillations so far not observed for {\it weak} modulations
\cite{not}. If we neglect these oscillations, it is approximately
given by $\sigma_{xy}^{nd}\approx ne/B$.

Experimentally one usually measures the resistivity
$\rho_{\mu\nu}$. Using the expressions given at the end of Sec.
III for $\rho_{\mu\nu}$, $\sigma_{xy}^{nd}\approx ne/B$,  and the
results for $\sigma_{yy}$  and $\sigma_{xx}$, we show in  Fig. \ref{fig8}
the resistivities divided by the magnetic field
$\rho_{xx}/B$, for $\alpha=\alpha_0$ ( upper panel) and
$\alpha=3\alpha_0$ ( lower panel), as a function of the magnetic field for a system
with $n_e=3n_0$ and otherwise the same parameters as in Fig. \ref{fig7}.
For $\alpha=\alpha_0$ and in the low-field region, in which the SdH oscillations are
absent, a beating pattern of the Weiss oscillations is
clearly observed in the $\rho_{xx}$ curve. The $\rho_{yy}$ curve
exhibits a beating pattern only for the SdH oscillations since
they result only from  collisional current contributions and the
Weiss oscillations are very weak as the diffusive contributions to
$\rho_{yy}\propto \sigma_{xx}$ vanish. For $\alpha=3\alpha_0$ though the beating patterns change:
that of the Weiss oscillations, when discernible  in $\rho_{xx}$, becomes shorter
and that of the SdH oscillations in $\rho_{yy}$  disappears. For completeness it should be mentioned,
though not shown, that for $\alpha=0$ there are no beating patterns in either the Weiss \cite{pee}
or SdH \cite{wan} oscillations.

\begin{figure}[tpb]
\vspace{-3cm}
\includegraphics*[width=90mm]{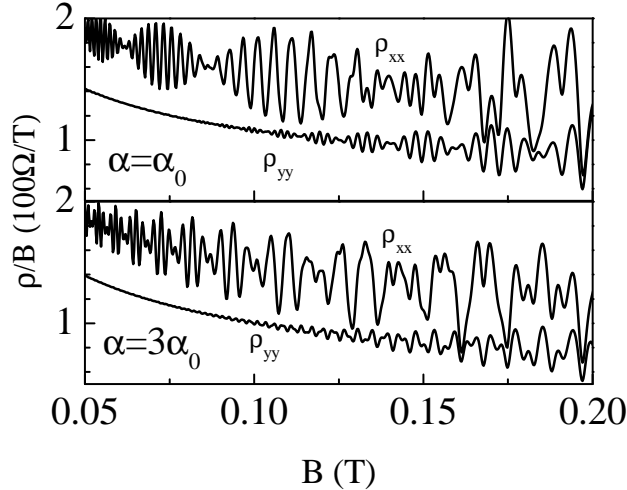}
\vspace{-2.5cm} \caption{Resistivities divided by the  field $B$, $\rho_{xx}/B$ and $\rho_{yy}/B$,
vs  field $B$ for $\alpha=\alpha_0$ ( upper panel) and
$\alpha=3\alpha_0$ ( lower panel). The other parameters are the same as in Fig. \ref{fig2}.}
\label{fig9}
\end{figure}

\section{Concluding remarks}

We evaluated {\it quantum mechanically} the dc conductivities of a 2DEG in the presence
of SOI of strength $\alpha$, of a normal magnetic field $B$, and of a {\it weak}
1D potential modulation of  strength $V_0$ and of period $a$. The SOI splits the Landau levels,
for $\alpha=0$, in two unequally spaced energy
branches. As in the absence of SOI,
the  modulation broadens the levels of these branches  into bands
and their bandwidths oscillate independently with the field $B$. This gives rise to two
flat-band conditions, instead of one  for $\alpha=0$, and to the beating patterns of the Weiss oscillations.
As for the  SdH oscillations, their beating patterns for weak $\alpha$ are nearly independent of the  modulation, at least as long as the latter is weak, and agree with those of Ref. 17 obtained in
the absence of  modulation. However, for strong $\alpha$ an additional structure is obtained
and the  SdH oscillations split in two, cf. Fig. \ref{fig6}. We also noticed the even-odd filling
factor transition in the SdH oscillations and explained  it with  the help
of the broadened DOS. A similar observation  was made in Ref. 22 for strong modulations
and was explained by the corresponding DOS.

Regarding the Weiss oscillations the results for the diffusive
conductivity agree, as expected, for the relevant weak magnetic
fields and  high quantum numbers $s$, with those of the
{\it classical} evaluation
of Ref. \cite{mag}. However,  the
results for the collisional conductivity could not be obtained by
a {\it classical} treatment and, to our knowlege, are new. It is
well known that this collisional or hopping conductivity describes
the SdH oscillations which cannot be treated  {\it classically}.
This explains their absence from Ref. \cite{mag} and their
modification for strong $\alpha$ as well as for strong $B$, cf.
Fig. \ref{fig6}, presented here.

For weak $\alpha$ both conductivities exhibit beating patterns. Those  of the diffusive
conductivity pertain to the Weiss oscillations and are due to the two independent  frequencies involved
in the bandwidths of the $+$ and $-$ branches created by the SOI whereas those of the
collisional conductivity pertain to the SdH oscillations 
and have  a similar explanation though the two frequencies involved here are not those of the bandwidths,
see the discussion of Fig. \ref{fig3}.  As we saw though, these  patterns weaken or
disappear rather quickly upon increasing the temperature or the strength $\alpha$.
On the electron density $n_e$
though, they appear to have a rather weak dependence, cf. Fig. \ref{fig7},
at least as long as $n_e$ falls in the range of
the usual experimental densities of a 2DEG.

We are not aware of any directly relevant experimental work. We hope though that the
findings described above will motivate experiments in which the magnetoresistivities along the $x$ and $y$ directions could be measured in a weakly modulated 2DEG in the presence of SOI. For a 1D modulation along the $x$ direction, the diffusive  and
collisional contributions to the conductivity can be obtained separately using the relations
$\sigma_{yy}=\sigma_{yy}^{dif}+\sigma_{yy}^{col}$ and  $\sigma_{xx}=\sigma_{xx}^{col}$. Combining them with
the standard relations given after Eq. (26), gives the magnetoresistivities.

\leftline{\bf Acknowledgments}

This work was supported by the Canadian NSERC Grant No. OGP0121756,
by the Flemish Science Foundation (FWO-VI),  the Belgian
Science Policy, and the EU-CERION project.

\end{document}